\documentclass[aps,prl,twocolumn,groupedaddress,amsmath]{revtex4}

\usepackage{bm} 

\usepackage{comment}
\usepackage{graphicx}
\usepackage{amssymb}

\usepackage{times} 
\usepackage{amsmath,amsfonts,latexsym}

\newcommand*{\cU}{{\cal U}}

\newcommand*{\cT}{{\cal T}}

\newcommand*{\cI}{{\cal I}}

\newcommand*{\pop}{\psi^{\vphantom{\dagger}}}
\newcommand*{\pdop}{\psi^\dagger}

\newcommand*{\phop}{\phi^{\vphantom{\dagger}}}

\newcommand*{\aop}{a^{\vphantom{\dagger}}}
\newcommand*{\adop}{a^\dagger}
\newcommand*{\bop}{b^{\vphantom{\dagger}}}
\newcommand*{\bdop}{b^\dagger}

\newcommand*{\sgn}{\mathrm{sgn}}

\newcommand{\vect}[1]{\mathbf{#1}}

\begin{document}

\title{Kondo polarons in a one-dimensional Fermi gas}

\author{Austen Lamacraft} 
\email{austen@virginia.edu}
\affiliation{Department of Physics, University of Virginia,
Charlottesville, VA 22904-4714 USA}
 
\date{\today}

\begin{abstract}
We consider the motion of a spin-1/2 impurity in a one-dimensional gas of spin-1/2 fermions. For antiferromagnetic interaction between the impurity and the fermions, the low temperature behavior of the system is governed by the two-channel Kondo effect, leading to the impurity becoming completely opaque to the spin excitations of the gas. As well as the known spectral signatures of the two-channel Kondo effect, we find that the low temperature mobility of the resulting `Kondo polaron' takes  the universal form $\mu\to \frac{3\hbar v_F^2}{2\pi k_B^2T^2}$, in sharp contrast to the spinless case where $\mu\propto T^{-4}$.
\end{abstract}

\maketitle

The motion of an impurity in a quantum liquid is one of the recurring paradigms of condensed matter physics. Variants of the problem appear in such well studied situations as the motion of ions in $^3\mathrm{He}$, muons and positrons in metals, and holes in metals and semiconductors~\cite{Rosch1999}. Certain multicomponent quantum liquids can also be studied in a dilute limit where the atoms of one component may be treated as individually interacting with a thermodynamically large number of atoms of the other components. Solutions of  $^3\mathrm{He}$ in  $^4\mathrm{He}$ provide the classic example of this situation~\cite{Bardeen1967}, while the field of ultracold atomic physics offers new possibilities~\cite{combescot:180402}.

In most theoretical treatments it is usual to ignore the dynamics of the impurity spin, if present. In this Letter we will show that for the case of a spin-1/2 impurity moving in a one-dimensional spin-1/2 Fermi gas the spin dynamics can completely change the low temperature behavior of the system. In fact this system manifests the two-channel Kondo (2CK) effect~\cite{nozieres1980}, in which a simple picture of scattering of fermions of the gas from the impurity is inapplicable at low temperatures. Instead the impurity becomes totally opaque to the collective spin excitations of the gas while being transparent to the density excitations. As well as unusual spectral properties associated with the 2CK effect, the low temperature mobility of the impurity  assumes the universal form $\mu\to  \frac{3\hbar v_F^2}{2\pi k_B^2T^2}$, where $v_F$ is the Fermi velocity of the gas. We will not be concerned with phenomena of the `X-ray edge' type, in which the impurity is added to or removed from the system, with the tunneling probability being affected by the orthogonality catastrophe~\cite{mahan1967,anderson1967,nozieres1969}.

The natural application of this model is to the dynamics of holes in doped semiconductor nanowires~\cite{calleja1991}, although there exists the possibility of using $^{171}\mathrm{Yb}$ as a spin-1/2 impurity in an ultracold quantum gas~\cite{kohl}. A realization of the two-channel Kondo effect without fine tuning of parameters is all the more striking given how elusive this remarkable phenomenon has proven, succumbing to experimental observation only last year~\cite{Potok2007},


In the present problem, this behavior is a consequence of the restricted scattering processes in one dimension. Low energy scattering events near the Fermi surface have momentum transfers close to a multiple of $2p_F$, where $p_F$ is the Fermi momentum. Thus for an impurity of mass $M$ there is a characteristic energy scale $E_{\mathrm{recoil}}\equiv 2p_F^2/M$ associated with a $2p_F$ momentum transfer. At temperatures $k_BT\ll E_{\mathrm{recoil}}$ such processes are frozen out, and the motion of the impurity is determined by forward scattering.

Consideration of the kinematics of these forward scattering events shows that processes involving a single fermion are also suppressed. Let us write the dispersion relation of the fermions as $\xi(p)=p^2/2m-\mu$ for chemical potential $\mu$, and that of the impurity as $\epsilon(p)=p^2/2M$. Scattering of a fermion with momentum $p\sim \pm p_F$ and low momentum transfer $q$ leads to a change in energy of $\xi(p+q)-\xi(p)\sim \pm v_F q$. Since the corresponding change in energy of the impurity is on the order of $q^2/2M$, energy conservation would require $q\sim 2Mv_F=2(M/m)p_F$, or energies $\sim (M/m)^2E_{\mathrm{recoil}}$.

These arguments suggest that the low temperature transport of the impurity is due to higher order processes. The simplest such process involves the scattering of two particles with momenta lying close to each fermi point. For the spinless case we use a model interaction $H_{\mathrm{int}}=V\sum_i\delta(x_i-X)$ between the fermions at positions $\{x_i\}$ and the impurity at $X$. The lowest order amplitude  for fermions with momenta $k_1$ and $k_2$ to scatter to $k_1+q_1$ and $k_2+q_2$, while the impurity momentum goes from $K$ to $K-q_1-q_2$ is $\cT^{(2)}\delta(E_i-E_f)$ 
with 
\begin{widetext}
\begin{eqnarray*}
\cT^{(2)}_{k_1,k_2,K\to k_1+q_1,k_2+q_2,K-q_1-q_2}&=&i\left(\frac{V}{L}\right)^2\left[\frac{1}{\xi_{k_1+q_1}-\xi_{k_1}+\epsilon_{K-q_1}-\epsilon_K}-\frac{1}{\xi_{k_2+q_2}-\xi_{k_1}+\epsilon_{K-k_2-q_2+k_1}-\epsilon_K}\right.\nonumber\\
&&\left.+\frac{1}{\xi_{k_2+q_2}-\xi_{k_2}+\epsilon_{K-q_2}-\epsilon_K}-\frac{1}{\xi_{k_1+q_1}-\xi_{k_2}+\epsilon_{K-k_1-q_1+k_2}-\epsilon_K}\right].
\end{eqnarray*}
\end{widetext}
At low $q_1$, $q_2$ the first and third terms give rise to singular behavior arising forward scattering processes at second order
$\cT^{(2)}\to i\left(\frac{V}{L}\right)^2\frac{q_2-q_1}{v_Fq_1q_2}$.
Despite this singularity the momentum relaxation rate of the impurity, given by 
\begin{eqnarray*}
\tau_\mathrm{mom}^{-1}=\frac{2\pi}{\hbar MT}\sum_{k_1,k_2,q_1,q_2}(q_1+q_2)^2|\cT^{(2)}|^2\delta(E_i-E_f) \\
\times n_{k_1}n_{k_2}\left(1-n_{k_1+q_1}\right)\left(1-n_{k_2+q_2}\right)
\end{eqnarray*}
($n_k$ is the Fermi distribution) is finite and vanishes as $T^{4}$, leading to a low temperature mobility $\mu=\tau_\mathrm{mom}/M\propto T^{-4}$~\cite{kagan1986epe,castro-neto1994,castroneto1996}. This calculation, together with the observation that even an almost opaque impurity will appear transparent at low temperatures as the backscattering processes are suppressed, offers a qualitative picture of behavior of the mobility from high to low temperatures in the spinless case~\cite{castroneto1996}.

%
\begin{figure}
\centering  \includegraphics[width=0.35\textwidth]{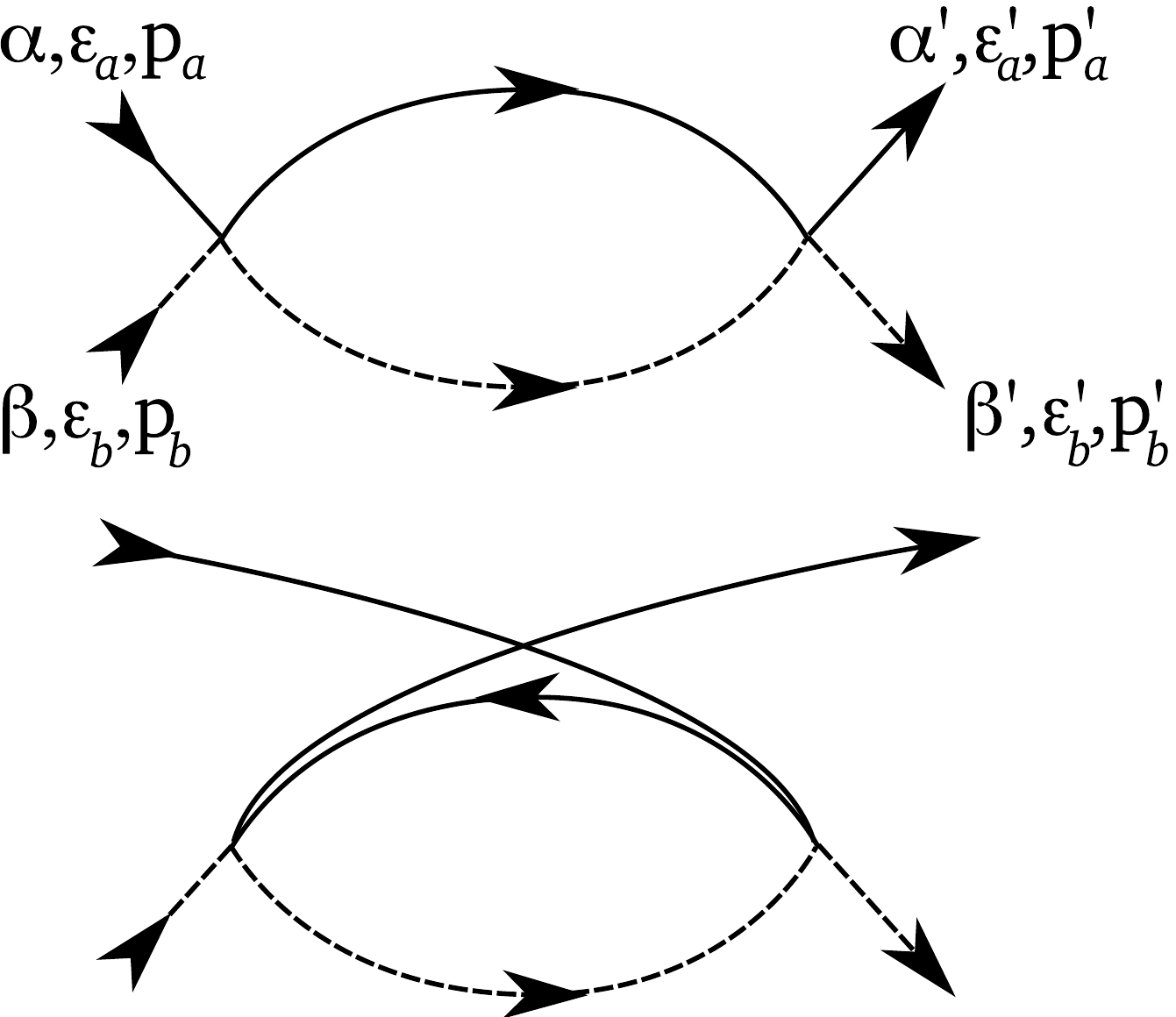}
\caption{Second order particle-particle (top) and particle-hole (bottom) contributions to the effective spin-spin interaction between the fermions (solid lines) and impurity (dashed lines)
 \label{fig:diagrams}}
\end{figure}

The validity of these results depended upon our being able to treat the interaction as weak at low temperatures. We will now show that when the impurity and the fermions both have spin-1/2 the most general form of $H_{\mathrm{int}}$ acquires singular contributions at low energies from higher order processes. The resulting divergences signal the need for  a radically different description at low temperatures.  It is convenient to use a second quantized representation of the Hamiltonian (setting $\hbar=k_B=1$)
\begin{eqnarray*} \label{2nd}
H_0&=&\sum_s\int dx \left[\frac{1}{2m}\partial_x\adop_{s}\partial_x \aop_{s}+\frac{1}{2M}\partial_x\bdop_{s}\partial_x \bop_{s}\right]\\\nonumber
H_{\mathrm{int}}&=&\int dx\,V\rho_a(x)\rho_b(x)+J\vect{S}_{a}(x)\cdot \vect{S}_b(x),
\end{eqnarray*}
Here $\rho_a(x)=\sum_\alpha\adop_\alpha(x)\aop_\alpha(x)$ and $\vect{S}_a(x)=\frac{1}{2}\sum_{\alpha,\alpha'}\adop_{\alpha}(x){\bm \sigma}_{\alpha\alpha'}\aop_{\alpha'}(x)$ denote the density and spin density of the fermions, and $\rho_b(x)$ and $\vect{S}_b(x)$ the corresponding quantities for the impurity. $V$ and $J$ parameterize the most general form of rotationally invariant interactions between the impurity and the fermions. Although it is important for the problem we discuss that the bare potential interaction $V$ is sufficiently strongly repulsive to prevent the formation of a singlet bound state for $J>0$, its renormalized strength tends to zero at low energies as explained above, and we will neglect it in the following. 

Since we have only one $b$-particle, the bare Green's function is 
$G_{b}(\epsilon,p)=(\epsilon-\epsilon(p)+i\delta)^{-1}$ 
%
%
while for the $a$-particle we have 
$G_{a}(\epsilon,p)=(\epsilon-\xi(p)+i\delta\sgn(\epsilon))^{-1}$. Two diagrams contribute to the renormalization of $J$ by second order processes. The effective interaction vertex (see Fig.~\ref{fig:diagrams}) has the second order contribution
\begin{widetext}
\begin{eqnarray*}
\Gamma^{(2)}_{\mathrm{eff}}=-\frac{J^2}{8}\sum_{i,j}\left[(\sigma^i\sigma^j)_{\alpha\alpha'}(\sigma^i\sigma^j)_{\beta\beta'}\cI_{pp}(\varepsilon_a+\varepsilon_b,p_a+p_b)+(\sigma^j\sigma^i)_{\alpha\alpha'}(\sigma^i\sigma^j)_{\beta\beta'}\cI_{ph}(\varepsilon_{a}'-\varepsilon_{b},p_{a}'-p_{b})\right],
\end{eqnarray*}
\end{widetext}
where the first and second terms are the contribution of the upper and lower diagrams respectively in Fig.~\ref{fig:diagrams}.
We have the explicit expressions.
\begin{eqnarray}\label{pp}
\cI_{pp/ph}(\omega, q)&\equiv&-i\int \frac{d\varepsilon}{2\pi}\frac{dp}{2\pi} G_{a}(\varepsilon,p)G_{b}(\pm\omega\mp\varepsilon,\pm q\mp p)\nonumber\\
&=&\int \frac{dp}{2\pi}\frac{\theta(\pm\xi(p))}{\xi(p)\pm\epsilon(q-p)-\omega}.
\end{eqnarray}
Note that these have logarithmic singularities 
$\cI_{pp/ph}(\omega\to 0,q\to\pm p_F)\to \mp\frac{\nu}{2} \ln \left|\frac{(M\pm m)\omega}{2M\mu}+\cdots\right|$
%
, where the dots denote terms of second and higher order in $\omega$ and $q\mp p_F$, and $\nu=m_a/\pi p_F$ is the Fermi surface density of states. The singularities originate in the vanishing of the impurity dispersion $\epsilon(q-p)$ in the denominator in Eq.~(\ref{pp}). In dimension greater than one this occurs only on isolated points or lines on the Fermi surface, so singularities are absent from the corresponding integrals.
Using the commutation relations of the Pauli matrices, we find that the amplitude for elastic scattering from the impurity has a contribution 
\begin{equation}\label{gamma2_sing}
\Gamma^{(2)}_{\mathrm{eff}}\sim\frac{J^2\nu}{8}\sigma^i_{\alpha\alpha'}\sigma^i_{\beta\beta'}\ln \mu/|\varepsilon|
\end{equation}
where the energy $\varepsilon$ is measured from the Fermi surface. Eq.~(\ref{gamma2_sing}) represents a singular renormalization of the coupling $J$, and a breakdown of the weak coupling picture.

Such a divergence is familiar from the perturbative treatment of the Kondo problem, which corresponds to the limit of the present model in which the mass of the impurity becomes infinite. The appearance of the impurity dispersion in the denominators of Eq.~(\ref{pp}) does not alter the low-energy singularity at leading order except to remove the contribution due to backscattering from the impurity, which is gapped by the energy $E_{\mathrm{recoil}}$. This leads to Eq.~(\ref{gamma2_sing}) being smaller by a factor of two than in the infinite mass case. 

There is, however, a more fundamental difference between the two cases: the absence of backscattering at low energies means that the left and right moving fermions form two distinct \emph{channels} in which the coupling $J$ becomes strong. For infinite mass, by contrast, only the even fermion modes (defined relative to the position of the impurity) are coupled to the impurity, with the odd mode decoupling entirely, giving a single channel. The low temperature behavior of the two-channel Kondo problem is completely different from the single-channel case. While the latter has an effective Fermi liquid description, the former cannot be described simply in terms of fermionic quasiparticles carrying charge and spin. The low temperature mobility of the impurity spin is therefore determined by a completely different mechanism than an impurity without spin. 

To establish these facts and to describe the physics of this model beyond simple perturbation theory, it is convenient to pass to a bosonized representation of the fermions. First we present the Hamiltonian in the mixed form
\begin{eqnarray*}\label{1D_H}
H=\sum_{s=\uparrow,\downarrow}\int dx \,\frac{1}{2m}\partial_x\adop_{s}\partial_x \aop_{s}+\frac{P^2}{2M}
+J\vect{S}_{a}(X)\cdot \vect{S},
\end{eqnarray*}
with $\left[P,X\right]=-i$. Next it is useful to pass to a frame co-moving with the impurity atom via the  transformation $\cU=e^{iX P_a}$, where $P_a$ is the total momentum of the $a$ particles. The effect of this transformation is
%
$\aop_{s}(x)\to \aop_{s}(x-X)$, 
$P\to P-P_a$ ,
%
thus eliminating the $b$ coordinate from the interaction, at the expense of introducing the momentum of the $a$ particles into the impurity kinetic energy.
\begin{eqnarray}\label{1D_H_comoving}
H=\sum_{s=\uparrow,\downarrow}\int dx \,\frac{1}{2m}\partial_x\adop_{s}\partial_x \aop_{s}+\frac{(P-P_a)^2}{2M}
+J\vect{S}_{a}(0)\cdot \vect{S}.
\end{eqnarray}
Note that after this transformation the variable $P$ is conserved due to the absence of $X$ from the resulting Hamiltonian, and simply corresponds to the total momentum of the system. We set $P=0$ from now on. It is convenient to express Eq.~(\ref{1D_H_comoving}) in a boson representation where $a_s(x)\sim\eta_{R,s}e^{i(p_Fx+
\phop_{R,s}(x))}+\eta_{L,s}e^{-i(p_Fx+
\phop_{L,s}(x))}$, with the fields $\phop_{L/R,s}(x)$ parameterizing the fluctuating Fermi sea near the two Fermi points~\cite{Haldane1981}. The anticommuting variables $\eta_{L/R,s}$ will be discussed shortly; first note that the anticommutation relations for the $a_{L/R,s}(x)$ are reproduced if we write the mode expansion for $\phop_{R/L,s}(x)$ (taking periodic boundary conditions on a system of size $L$)
\begin{eqnarray}
\phop_{R/L,s}(x)&=&\phi_{R/L,s}^{(0)}+\frac{2\pi x}{L}N_{R/L,s}\nonumber\\
&&+\sum_{n=1}^\infty\frac{1}{\sqrt{n}}\left(c_{R/L,s}^{(n)}e^{\pm iq_nx}+\mathrm{h.c.}\right),
\end{eqnarray}
where $q_n=2\pi n/L$. The mode operators $c^{(n)}_{R/L,s}$ and their conjugates satisfy the canonical Bose commutation relations, while the zero modes $\phi^{(0)}$ and $N_{R/L,s}$ satisfy $\left[\phi_{R/L,s}^{(0)},N_{R/L,s}\right]=\mp i$. We therefore have
\begin{equation*}\label{fundamental}
\left[\phop_{L/R,s}(x),\partial_{x'}\phop_{L/R,s'}(x')\right]=\pm 2\pi i\delta_{ss'}\delta(x-x').
\end{equation*}
%
%
%

In the following two technical points are important: i) the possibility of nonzero winding $\phop_{R/L,s}(L)-\phop_{R/L,s}(0)=2
\pi N_{R/L,s}$  which describes uniform shifts of the Fermi points when a particle is added or removed from the corresponding branch (as occurs in backward scattering) and ii) the careful treatment of the operators $\eta_{p,s}$ that satisfy $\{\eta_{p,s},\eta_{p',s'}\}=2\delta_{pp'}\delta_{ss'}$. The bosonized Hamiltonian involves only bilinears of these operators. The operator $\eta_{L,\uparrow}\eta_{L,\downarrow}\eta_{R,\uparrow}\eta_{R,\downarrow}$ commutes with all these bilinears, so we may set it equal to $\pm 1$, which gives relations between bilinears e.g. $\eta_{L,\uparrow}\eta_{L,\downarrow}=\mp\eta_{R,\uparrow}\eta_{R,\downarrow}$.

The total Hamiltonian has the form $H=H_K+H_{\mathrm{fs}}+H_{\mathrm{bs}}$, where the kinetic part $H_K$ is
\begin{equation*}\label{kinetic}
H_K=\frac{v_F}{4\pi}\sum_{\substack{p=R,L\\ s=\uparrow,\downarrow}} \int dx \left(\partial_x\phop_{p,s}\right)^2+\frac{p_F^2}{2M}\left[\sum_s N_{R,s}-N_{L,s}\right]^2.
\end{equation*}
The terms $H_{\mathrm{fs}}$ and $H_{\mathrm{bs}}$ describe the forward scattering and backward scattering parts of the original interaction. For the moment we will drop $H_{\mathrm{bs}}$, reintroducing it later. Allowing for the possibility of anisotropic couplings, we have for $H_{\mathrm{fs}}$
\begin{multline}\label{H_fs}
H_{\mathrm{fs}}=\frac{J_{0\perp}}{\lambda}\left[\eta_{L,\uparrow}\eta_{L,\downarrow}e^{i\phop_X(0)}+\eta_{R,\uparrow}\eta_{R,\downarrow}e^{-i\phop_X(0)}\right]e^{i\phop_S(0)}S^-\\+\mathrm{h.c.}+J_{0\parallel}\partial_x\phop_S(0)S^z
\end{multline}
%
where $\lambda$ is a short-distance cutoff and $J_{0\perp}$ and $J_{0\parallel}$ are the transverse and longitudinal parts of the spin-spin interaction. Eq.~(\ref{H_fs}) is written in terms of the chiral fields
\begin{equation*}
\left(\begin{array}{c}\phop_C(x) \\\phop_S(x) \\\phop_F(x) \\\phop_X(x)\end{array}\right)=\frac{1}{2}\left(\begin{array}{cccc}1 & -1 & 1 & -1 \\1 & -1 & -1 & 1 \\1 & 1 & 1 & 1 \\1 & 1 & -1 & -1\end{array}\right)\left(\begin{array}{c}\phop_{L,\uparrow}(x) \\\phop_{R\uparrow}(-x) \\\phop_{L,\downarrow}(x) \\\phop_{R,\downarrow}(-x)\end{array}\right)
\end{equation*}
%
%
We can now apply the Emery-Kivelson transformation $\cU_{\mathrm{EK}}H\cU_{\mathrm{EK}}^{\dagger}$ with $\cU_{\mathrm{EK}}=\exp(iS^z\phop_S)$~\cite{emery1992} . This removes the $e^{\pm i\phop_S}$ factors from $H_{\mathrm{fs}}$, while shifting the kinetic energy by $-v_FS^z\partial_x\phop_X$. Thus for the special value $J_{0\parallel}=v_F$, the Hamiltonian takes the form 
\begin{eqnarray} \label{boson_fs}
H&=&\frac{v_F}{4\pi}\sum_{p=C,F,S,X} \int dx \left(\partial_x\phop_{p}\right)^2\nonumber\\
&&+\frac{2iJ_{0\perp}}{\lambda}\eta_{L,\uparrow}\eta_{L,\downarrow}S^x\sin\phi_X(0) +E_{\mathrm{recoil}}N_F^2
\end{eqnarray}
The impurity term in Eq.~(\ref{boson_fs}) can be expressed in terms of a fermion: $
\sin\phi_X\sim \pop_X+\pdop_X$. Although the resulting Hamiltonian may be solved exactly, we need only observe that simple scaling implies that the impurity term dominates at low energy, leading to the boundary condition $\phi_X(0_+)=-\phi_X(0_-)$ (recall that $\phi_X$ is a chiral field and satisfies a first order wave equation)~\cite{maldacena1997}. This conclusion is not altered by for $J_{0\parallel}$ different from $v_F$ (providing the interactions remain antiferromagnetic in sign): the perturbing operator $S^z\partial\phi_X(0)$ is irrelevant at the low energy fixed point. 

Although the 2CK boundary condition cannot be understood easily in terms of the fermions, its physical meaning is simple. It corresponds to the perfect reflection of the spin mode at the impurity: $\phop_{L,\uparrow}-\phop_{L\downarrow}|_{0_\pm}=\phop_{R,\uparrow}-\phop_{R\downarrow}|_{0_\pm}$, while the density mode propagates unhindered. This dramatic manifestation of spin-charge separation in the 2CK effect also determines the low temperature mobility of the resulting `Kondo polaron', as we discuss below.

The crucial difference between the cases of finite and infinite impurity mass arises from the effect of $H_\mathrm{bs}$, which after the transformation $\cU_\mathrm{EK}H_\mathrm{bs}\cU_\mathrm{EK}^\dagger$ takes the following form at the fixed point.
\begin{equation*}
H_{\mathrm{bs}}\to-\frac{2iJ_{2k_F\perp}}{\lambda}\eta_{L,\uparrow}\eta_{R,\downarrow}S^y\cos \phop_{F}(0)
\end{equation*}
%
In the absence of an impurity kinetic energy term this is a relevant perturbation of dimension $1/2$ that leads to the one-channel Kondo fixed point. Indeed for $J_{0\perp}=J_{2k_F\perp}$, $H+H_{\mathrm{bs}}$ can be written in terms of the resonant level model characteristic of the Toulouse point~\cite{Toulouse1970}. The boundary conditions that result when both $J_{0\perp}$ and $J_{2k_F\perp}$ flow to strong coupling are $\phop_{X}(0_+)=-\phop_{X}(0_-)$ and $\phop_{F}(0_+)=-\phop_{F}(0_-)\pm2\pi S^z$. This corresponds to $\aop_{L,s}(0_\pm)=-\aop_{R,s}(0_{\pm})$ (accounting for the phase shift in $\phop_S$ induced by the EK transformation): total reflection of fermions from the impurity with phase shift $\pi/2$ in the even channel (although this conclusion will be changed by the inclusion of non-zero potential scattering $V$)

$H_{\mathrm{bs}}$ changes the value of $N_F$ by $\pm 1$, however, so that this breakdown of the 2CK fixed point is prevented for finite impurity mass due to the recoil energy in Eq.~(\ref{boson_fs}). The effect of backscattering at low energies is described by an effective Hamiltonian obtained at second order 
\[H_{\mathrm{eff}}=-\frac{J_{2k_F,\perp}^2}{E_{\mathrm{recoil}}^2a}(\partial_x\phop_F(0))^2,\]
an irrelevant operator of dimension 2.

Although they have no charge, the spin excitations do carry momentum and exert a force on the impurity. In the low temperature limit, the recoil of the impurity in this process can be neglected, so the drag force due to spin excitations can be found by considering an impurity with constant velocity. The velocity dependence of the force due to the `Doppler shift' of the spin excitations gives at lowest order a dissipative force  $F_T=-(2P_T/v_F^2)\dot X$, if $P_T$ is the total incident power, and where $v_F$ plays the role of the speed of light~\cite{braginsky1967,jaekel1993}.  
The one-dimensional version of the Stefan-Boltzman law gives $P_T=2\times (\pi T^2/6)$ (the factor of 2 being for the two sides of the impurity), resulting in a mobility (with all dimensionful constants restored)
\begin{equation*}
\mu\to \frac{3\hbar v_F^2}{2\pi k_B^2T^2},\qquad\mathrm{as}\; T\to 0
\end{equation*}
Compared to the spinless case we see that the divergence of the mobility is much slower and furthermore \emph{universal}, being a characteristic of the 2CK fixed point rather than arising from some irrelevant perturbation of non-universal amplitude. At zero temperature the damping force experienced by the impurity is $-\frac{\hbar}{6\pi v_F^2}\dddot X$~\cite{Fulling1976}

The influence of the 2CK effect is also apparent in the spin fluctuations of the impurity, which has imaginary susceptibility $\mathrm{Im}\,\chi_{\mathrm{imp}}(\omega)=\frac{1}{2}\tanh\left(\omega/T\right)\Gamma/\left(\omega^2+\Gamma^2\right)$, with $\Gamma=J_{0\perp}^2/a$~\cite{emery1992}. At zero temperature the spin fluctuations extend to zero frequency, with an accompanying logarithmic divergence of the real susceptibility. A finite magnetic field splitting the degeneracy of the impurity spin states by a Zeeman energy $\Delta_Z$ restores the Fermi liquid behavior and the $T^{-4}$ behavior of the mobility at temperatures  $\lesssim\Delta_Z^2/\Gamma$~\cite{zarand2000}.

In conclusion, we have shown that the motion of a spin-1/2 impurity in a one-dimensional spin-1/2 Fermi gas provides a realization of the two-channel Kondo effect without fine tuning. The unusual nature of the low temperature state formed by the coupling of the impurity and fermion spins gives rise to universal behavior of the mobility.

\end{document}